\documentclass[11pt]{article}
\pdfoutput=1

\usepackage{jheppub} 
\usepackage{amsmath,amssymb,epsfig,amsfonts}
\usepackage{epsf}
\usepackage{makeidx}
\usepackage{slashed}
\usepackage{verbatim} %for comments
\usepackage{float}
\restylefloat{table}
\usepackage{pdflscape}
\usepackage{tikz}
\usepackage{pifont}
\usepackage{array}
\usepackage{youngtab}
\usepackage{multirow}
\usepackage{xcolor}
\usepackage{ulem}
\usepackage{cleveref}
\usepackage{tensor}
\usepackage{todonotes}
\usepackage{mathrsfs}
\usepackage{changepage}
\usepackage{anyfontsize}
\usepackage{caption}
\makeatletter

\newcommand{\be}{\begin{equation}}
\newcommand{\ee}{\end{equation}}
\newcommand{\ba}{\begin{aligned}}
\newcommand{\ea}{\end{aligned}}

\newcommand{\dd}{\mathrm{d}}

\newcommand{\me}{\mathrm{e}}
\newcommand{\ii}{\mathrm{i}}
\newcommand{\vol}{\mathrm{vol}}

\makeatother

\begin{document}

% format
\baselineskip=18pt  % a la harvmac
\numberwithin{equation}{section}  % make eq labels (sec.num)
\allowdisplaybreaks  % allow page breaks in displayed eqs

%%%%%%%%%%%%%%%%%%%%%%%%%%%%%%%%%%%%%%%%%%%
%%%        TITLE BEGINS HERE
%%%%%%%%%%%%%%%%%%%%%%%%%%%%%%%%%%%%%%%%%%%

%% ========== title (note version) begins here ==========
%
%\vspace*{-1cm}
%\begin{center}
% {\Large\bf Title of the Document}
%\end{center}
%\vspace*{-.5cm}
%
%% ========== title (note version) ends here ==========

%% ========== title (paper version, a la harvmac) begins here ==========

\thispagestyle{empty}

% Report number
%\vspace*{-2cm} 
%\begin{flushright}
%{\tt KCL-MTH-18-NN}\\
%\end{flushright}

% title, authors, affiliation
\vspace*{1cm} 
\begin{center}
{\LARGE  A tale of (M)2 twists} 
 \vspace*{1.8cm}
 
 \renewcommand{\thefootnote}{}
\begin{center}
 {Christopher Couzens
 \footnotetext{cacouzens@khu.ac.kr}}
\end{center}
\vskip .2cm
%
%
% \vspace*{.5cm} 
{ Department of Physics and Research institute of Basic Science, \\
  Kyung Hee University, Seoul 02447, Republic of Korea }\\
  
 {\tt {}}

\vspace*{0.8cm}
\end{center}

 \renewcommand{\thefootnote}{\arabic{footnote}}
 
 \begin{adjustwidth}{0.26in}{0.26in}

\begin{center} {\bf Abstract } \end{center}

\vspace*{1cm}
\noindent

We study the parameter space of magnetically charged AdS$_2\times\mathbb{WCP}^{1}_{[n_-,n_+]}$ solutions in 4d U$(1)^4$ gauged STU supergravity. We show that both \emph{twist} and \emph{anti-twist} solutions are realised and give constraints for their existence in terms of the magnetic charges of the solution. We provide infinite families of both classes of solution in terms of their magnetic charges and weights of the orbifold. As a byproduct of our analysis we obtain a closed form expression for the free-energy of the 4-charge magnetic solution in terms of the magnetic charges and weights $n_{\pm}$. We also show that the AdS$_2$ solution is the near-horizon of an asymptotically AdS$_4$ black hole which can be found in the literature.

\noindent 
\end{adjustwidth}

%\vspace*{.5cm}

% abstract
%\noindent

\newpage
%%%%%%%%%%%%%%%%%%%%%%%%%%%%%%%%%%%%%%%%%%%
%%%           TITLE ENDS HERE
%%%%%%%%%%%%%%%%%%%%%%%%%%%%%%%%%%%%%%%%%%%

\tableofcontents
\printindex

%%%%%%%%%%%%%%%%%%%%%%%%%%%%%%%%%%%%%%%%
%%%        MAIN TEXT BEGINS HERE
%%%%%%%%%%%%%%%%%%%%%%%%%%%%%%%%%%%%%%%%%%%

\section{Introduction}

There has been a recent surge in studying supersymmetric AdS$_{d-2}\times\Sigma$ solutions of $d=4,5,6,7$ gauged supergravity, where $\Sigma$ is a Riemann surface admitting a non-constant curvature metric. Solutions where $\Sigma$ is the weighted projective space $\mathbb{WCP}^{1}_{n_+,n_-}$, also known as a spindle, were first found in \cite{Ferrero:2020laf}, and later extended in \cite{Ferrero:2020twa,Hosseini:2021fge,Faedo:2021nub,Couzens:2021rlk,Ferrero:2021ovq,Faedo:2021kur,Ferrero:2021etw,Cassani:2021dwa,Boido:2021szx}. These solutions are naturally interpreted as arising from compactifying M2-, D3-, D4-, and M5-branes on the spindle and can be uplifted to 10- or 11-dimensional supergravity. One particularly interesting aspect of these solutions is the manner in which they preserve supersymmetry. The AdS$_3\times \Sigma$ solutions were argued to be dual to the compactification of certain $\mathcal{N}=1$ SCFTs on the spindle, \cite{Ferrero:2020laf}, where supersymmetry is preserved by a new mechanism dubbed the \emph{anti-twist}. Contrary to the more canonical topological twist, the background R-symmetry vector is not identified with the spin-connection of the spindle for the anti-twist and the spinors are non-constant. It was later shown in \cite{Ferrero:2020twa} that the same mechanism was in play for the M2-brane geometries in 4d Einstein--Maxwell and later in \cite{Ferrero:2020twa,Ferrero:2021ovq} for the $X^0X^1$ truncation. For the M5-brane and D4-brane geometries, supersymmetry is preserved by yet another mechanism, \cite{Ferrero:2021wvk,Faedo:2021nub} dubbed a topological topological twist, or \emph{twist} from now on. As in the usual topological twist the charge of the field strength of a background R-symmetry vector through the spindle is the Euler character of the spindle, however the spinor is not globally a constant as in the usual topological twist.\footnote{The recent paper \cite{Ferrero:2021etw} which appeared shortly before submission, shows that the twist and anti-twist are the only possible ways of realising supersymmetry on a spindle.}

Another interesting class of Riemann surface with non-constant curvature, which were initially developed in parallel, are the topological discs of \cite{Bah:2021mzw,Bah:2021hei}. The Riemann surface $\Sigma$ has the topology of a disc, with the boundary a smeared M5 brane, and are holographic duals of Argyres--Douglas theories. Topological discs were later found for M2-, D3- and D4- branes in $d=4,5,6$ gauged supergravity in \cite{Couzens:2021tnv,Couzens:2021rlk,Suh:2021ifj,Suh:2021aik,Suh:2021hef}. It was noted in \cite{Couzens:2021tnv,Couzens:2021rlk} that the discs and spindle solutions are different global completions of the same local solutions, with the discs at a seemingly degenerate limit. In the uplifted theory, disc solutions are singular due to the presence of a smeared brane which wraps AdS$_{d-2}$ and the boundary of the disc. Similar mechanisms for preserving supersymmetry are in play for the discs and the spindles. 

In this short note we will further study the multi-charge M2-brane spindle solutions discussed in \cite{Couzens:2021rlk, Ferrero:2021ovq}. As noted in \cite{Couzens:2021rlk}, and also more recently in \cite{Ferrero:2021etw}, there is a possibility to realise solutions with both a \emph{twist} and \emph{anti-twist}. We will investigate the conditions that the magnetic charges of the solution need to satisfy in order for this to occur, and confirm that both twists are realised. We construct infinite families of solutions for given magnetic charges for both types of twist and we provide an expression for the free-energy of the two cases in terms of the four independent magnetic charges. We find that the free-energy takes a subtle yet different, form for the twist and anti-twist solutions, 
\be
\mathcal{F}=\frac{2\pi}{3n_+ n_-}N^{3/2}\sqrt{\hat{P}^{(2)}-\sigma n_+ n_- +\sigma \sqrt{\big(\hat{P}^{(2)}-\sigma n_+ n_-\big)^2 - 4 \hat{P}^{(4)}}}\, ,
\ee
with $\sigma=1$ for twist and $\sigma=-1$ for anti-twist. The $\hat{P}^{(a)}$ are the unique symmetric polynomials of power $a$ of the integer magnetic charges and are defined later.

This note is organised as follows. In section \ref{sec:review} we quickly review the AdS$_2\times \mathbb{WCP}^{1}_{[n_+,n_]}$ solutions of 4d U$(1)^4$ gauged supergravity. In section \ref{sec:intomagnetic} we give a thorough explanation for inverting the roots of the quartic which governs the solution in terms of the magnetic charges and orbifold weights. As a byproduct of this analysis we can provide a closed form expression for the free-energy in the multi-charge case. In section \ref{sec:antitwist} and section \ref{sec:twist} we use the expressions we derive for the roots in terms of the magnetic charges to construct infinite families of solutions of both types. In the first of two appendices we show that the solution is supersymmetric by explicitly computing the Killing spinors of the solution. We also briefly discuss the differences between the two twists at the level of the Killing spinors. In the second and final appendix we show that the AdS$_2$ solution we consider is the near-horizon of the asymptotically AdS$_4$ solution found in \cite{Lu:2014sza}.

{\bf Note added:}
Whilst writing up, \cite{Ferrero:2021etw} appeared on the arXiv which has overlap with this paper. Amongst other interesting aspects of their work, they give the roots for a single twist solution, proving the existence but are not able to give the integer magnetic charges. From our construction we are able to do this and enlarge the known solutions to infinite families.

\section{A short review of AdS$_2\times \mathbb{WCP}^1_{[n_+,n_-]}$ solutions from wrapped M2 branes}\label{sec:review}

We will consider static AdS$_2$ solutions of 4d U$(1)^4$ gauged STU supergravity without axions which can be obtained as a consistent truncation of 11d supergravity on $S^7$.\footnote{The most general 4d U$(1)^4$ gauged supergravity from a truncation of 11d supergravity on $S^7$ contains in addition 3 axions, however these may be consistently truncated out of the theory by requiring that $F^{I}\wedge F^{J}=0$. The solution we consider satisfies this property and therefore it is consistent to set the axions to vanish. Generalisations of the solution to rotate and have non-trivial axions can be found in \cite{Ferrero:2021ovq,CCKStoappear}. } The bosonic field content of the truncation consists of a metric, four abelian gauge fields and four real scalars subject to a constraint. The action for the theory is
\begin{align}
S=\frac{1}{16 \pi G_{(4)}} \int \bigg( R-\frac{1}{2} \sum_{I=1}^{4} (X^{(I)})^{-2} \big(\dd X^{(I)}\big)^2 + \sum_{I<J} X^{(I)}X^{(J)}-\frac{1}{2} \sum_{I} \big(X^{(I)}\big)^{-2} \big|F^{I}\big|^2\bigg)\dd \vol_4\, ,\label{eq:action}
\end{align}
with the four scalars $X^{(I)}$ subject to the constraint $X^{(1)}X^{(2)}X^{(3)}X^{(4)}=1$. One may obtain the above Lagrangian from the general form of 4d $\mathcal{N}=2$ gauged supergravity coupled to three vector multiplets with prepotential 
\be
F=-\ii \sqrt{X^{(1)}X^{(2)}X^{(3)}X^{(4)}}\, .
\ee

The AdS$_2$ solution of 4d U$(1)^4$ gauged STU supergravity that we are interested in was originally found in 11d supergravity on a squashed $S^7$ in \cite{Gauntlett:2006ns} using a double Wick rotation of the 4d black hole solutions with spherical horizon studied in \cite{Cvetic:1999xp}. We will be interested in the 4d solutions obtained by truncating the theory on the $S^7$ as studied in \cite{Couzens:2021rlk}, see also \cite{Ferrero:2021ovq}. The solution is the near-horizon of an asymptotically AdS$_4$ magnetically charged black hole as we show in appendix \ref{app:fullblackhole}. The near-horizon geometry is
\begin{align}
\dd s^2 &=\sqrt{P(w)} \Big[ \dd s^2(\text{AdS}_2) +\dd s^2(\Sigma)\Big]\, ,\label{eq:NHmetric}\\
 \dd s^2(\Sigma)&=\frac{\dd w^2}{f(w)}+\frac{f(w)}{P(w)}\dd z^2\, ,\\
A^{I}&= -\frac{w}{2(w-c^I)}\dd z\, ,\\
X^{(I)}&= \frac{P(w)^{\frac{1}{4}}}{w-c^I}\, ,\label{eq:NHscalars}
\end{align}
with $f(w)$ and $P(w)$ quartic polynomials given by
\be\label{eq:fdef}
P(w)=\prod_{I=1}^{4} (w-c^I)\, ,\qquad f(w)= P(w)-w^2\, ,
\ee
and the solution depends on four free parameters, $c^I$. This may be uplifted to 11d supergravity on a seven-sphere and the resultant compact 9-dimensional space is a GK geometry \cite{Gauntlett:2007ts,Kim:2006qu,MacConamhna:2006nb}.

To bound the space we require $f(w)$ to admit two roots which define the domain of the line-interval parametrised by the coordinate $w$. Between the two roots we require that $f(w)$ is positive, this immediately implies that $P(w)$ is also strictly non-zero between two such roots. We will focus on the regularity condition where $f(w)$ admits single roots and $P(w)$ is strictly non-zero.\footnote{When $f(w)$ and $P(w)$ have a common root one obtains a topological disc solution as noted in \cite{Couzens:2021tnv,Couzens:2021rlk}, see also \cite{Suh:2021aik,Suh:2021hef,Suh:2021ifj,Bah:2021mzw,Bah:2021hei} for other disc solutions.} Since $f(w)$ is quartic and tends to infinity as $w\rightarrow \pm \infty$ it follows that for a well-defined region where $f(w)$ is positive we require the existence of four real roots, see for example figure 1 in \cite{Couzens:2021rlk}. Let us denote these roots by $w_1<w_2<w_3<w_4$. We are lead to bound the line interval between $[w_2,w_3]$ between which the metric has the correct signature.
Near such a single root the metric on $\Sigma$ becomes
\be
\dd s^2(\Sigma)\sim \frac{4}{|f'(w_*)|}\bigg[ \dd R^2 +\frac{|f'(w_*)|^2}{4 w_*^2}R^2 \dd z^2\bigg]\, ,
\ee
where we have defined the coordinate $R^2= |w-w_*|$. This is locally $\mathbb{R}^2/\mathbb{Z}_k$ if the periodic coordinate $z$ has period 
\be
\frac{\Delta z}{4\pi}=\frac{|w_*|}{k|f'(w_*)|}\, .
\ee
In the following we will allow the metric to have conical singularities at both end-points of the line-interval, fixing the period as
\be
\frac{\Delta z}{4\pi}=\frac{|w_2|}{n_- |f'(w_2)|}=\frac{|w_3|}{n_+ |f'(w_3)|}\, ,\label{eq:period}
\ee
with $n_{\pm}$ relatively prime integers which parametrise the conical deficit angles at the two poles, that is, there is a deficit angle of $2\pi (1-n_{\pm}^{-1})$ at the two poles. 

The magnetic charges of the solution are quantised as
\be\label{eq:Qdef}
Q^{I}=\frac{1}{2\pi}\int_{\Sigma} \dd A^I=\frac{\Delta z}{4\pi}\frac{(w_3-w_2)c^I}{(w_3-c^I)(w_2-c^I)}= \frac{p^{I}}{n_+ n_-}\, ,
\ee
with $p^{I}\in \mathbb{Z}$. With this quantisation, and the requirement that the $p^{I}$ are relatively prime to both $n_+$ and $n_-$ the 11d uplift on a seven-sphere is smooth, see \cite{Ferrero:2020twa,Ferrero:2021ovq,Couzens:2021rlk}.
The graviphoton of the gravity multiplet is identified to be the sum of the four gauge fields $A^{I}$, 
\be
A^R=\sum_{I=1}^{4} A^{I}\, ,
\ee
with the remaining independent combinations dual to flavour symmetries generically. One can compute the magnetic flux of the graviphoton through the spindle. As derived in \cite{Couzens:2021rlk} it takes the form
\begin{align}
Q^R=\sum_{I=1}^{4} Q^I =\frac{\Delta z}{4\pi} \bigg(\text{sgn}(w_3)\frac{|f'(w_3)|}{|w_3|}+\text{sgn}(w_2)\frac{|f'(w_2)|}{|w_2|}\bigg)\, ,
\end{align}
whilst a similar expression for the Euler character is 
\begin{align}
\chi=\frac{\Delta z}{4\pi}\bigg( \frac{|f'(w_3)|}{|w_3|}+\frac{|f'(w_2)|}{|w_2|}\bigg)\, .
\end{align}
Using the period as defined in \eqref{eq:period} we may write these as\footnote{These agree with the expressions derived recently in \cite{Ferrero:2021etw} using a similar computation.}
\begin{align}
\chi&=\frac{1}{n_+}+\frac{1}{n_-}\, ,\qquad
Q^R=\sum_{I=1}^{4} Q^I=\frac{\text{sgn}(w_3)}{n_+}+\frac{\text{sgn}(w_2)}{n_-}\, ,\label{eq:chisum}
\end{align}
with $\text{sgn}$ the usual sign function. Since $w\rightarrow -w,\,  c^{I}\rightarrow -c^{I}$ is a symmetry of the solution, we may always take the larger root to be positive. Without loss of generality we can take $\text{sgn}(w_3)=1$ and if we define $\text{sgn}(w_2)=\sigma=\pm 1$ we have\footnote{We have chosen $\sigma$ to match the conventions in \cite{Faedo:2021nub}.} 
\be
\chi=\frac{1}{n_+}+\frac{1}{n_-}\, ,\qquad Q^R=\frac{1}{n_+}+\frac{\sigma}{n_-}\, .
\ee
As observed in \cite{Couzens:2021rlk} and confirmed in \cite{Ferrero:2021etw} for roots of the same sign, $\sigma=1$ supersymmetry should be preserved by a \emph{topological topological twist}, whilst for roots of differing sign $\sigma=-1$ supersymmetry is not preserved by such a topological twist but instead by the so-called \emph{anti-twist} in the language of \cite{Faedo:2021nub,Ferrero:2021etw}.\footnote{The way supersymmetry is preserved for the different choices of twist has been nicely explained in \cite{Ferrero:2021etw} and to elucidate some of the points better we present the Killing spinors of our solution in appendix \ref{app:spinors} with some discussion about the two twists.} Solutions with $\sigma=-1$ were studied in \cite{Ferrero:2020twa} for the Einstein--Maxwell truncation and in \cite{Couzens:2021rlk,Ferrero:2021ovq} for the $X^{0}X^{1}$ truncation (pairwise equal gauge fields and constrained scalars). In \cite{Couzens:2021rlk} it was proven that only anti-twist solutions are possible in either of these truncations.

The motivation of this paper is to derive constraints on the magnetic charges and orbifold weights of the solution for preserving supersymmetry with either twist. From the above discussion this is equivalent to obtaining roots of the quartic $f(w)$ with the middle two roots either both positive (twist) or one positive and one negative (anti-twist). The basis of our analysis revolves around a clever rewriting that allows us to give closed form expressions for the roots, $w_I$ in terms of the magnetic charges of the solution and the orbifold weights $n_{\pm}$. To proceed it is useful to split the discussion into the two cases where $w_2<0<w_3$ realising the anti-twist or where both roots are positive realising the twist. Note that the both negative roots case $w_2<w_3<0$, can be obtained from the two positive roots case by sending $w\rightarrow -w$ and $c^I\rightarrow -c^I$. Therefore without loss of generality we take $w_3>0$. Note that the full mapping is:
\be\label{eq:symmetry}
(w, c^I, p^I, n_+,n_-)\longrightarrow (-w,-c^I,-p^I,n_-,n_+)\, ,
\ee
in particular it sends the magnetic charges to minus themselves and interchanges the orbifold weights.

Before we conclude this introductory solution let us present the large $N$ free-energy of the solution. It takes the compact expression
\be 
\mathcal{F}=\frac{1}{4 G_2}=\frac{2\sqrt{2}\pi }{3} N^{3/2} \frac{(w_3-w_2)\Delta z}{4\pi}\, .\label{eq:freeE}
\ee
This result is somewhat unsatisfactory since it is given in terms of unphysical parameters, as a byproduct of our analysis we will derive an expression in terms of only physical parameters; magnetic charges and the orbifold weights. 
Note that this structure of the free energy depending only on the period and difference between the two roots is also true for the analogous D3 brane solutions \cite{Ferrero:2020laf,Boido:2021szx}.

%%
%
%				Roots
%%%

\section{Determining the roots of the quartic}\label{sec:intomagnetic}

In this section we will perform some clever manipulations of the roots of the polynomial $f(w)$ which allows us to write the roots in terms of the charges and $n_{\pm}$ whilst also eliminating the constants $c_I$, in fact by the end of this section these will not appear again in this paper. 
As discussed above, and recalling that we take $w_3>0$ without loss of generality, the two types of twist are realised when
\be
\begin{cases}
w_2<0<w_3\, ,\qquad \text{anti-twist}\\
0<w_2<w_3\, ,\qquad \text{twist}
\end{cases}
\ee
To cover both cases let us use the parameter $\sigma=\pm1$ introduced earlier to write $\sigma w_2=|w_2|$.
Then equation \eqref{eq:chisum} reads
\be
\chi=\frac{1}{n_+}+\frac{1}{n_-}\, ,\qquad \sum_{I=1}^{4} Q^I=\frac{1}{n_+}+\frac{\sigma}{n_-}\, .
\ee
We now want to compute the roots in terms of the four charges $Q^I$ and the orbifold parameters $n_{\pm}$, however since the parameters $c^I$ appear in the charges $Q^I$ and the roots depend non-trivially on them this is somewhat complicated.
Instead it is useful to use the following symmetric combinations of the magnetic charges
\begin{alignat}{3}
\hat{Q}^{(1)}&\equiv\sum_{I=1}^{4} Q^{I}&&=\frac{1}{n_+n_-}\sum_{I=1}^{4} p^{I}&&\equiv\frac{1}{n_+ n_-}\hat{P}^{(1)}\, ,\nonumber\\
\hat{Q}^{(2)}&\equiv\sum_{1\leq I<J\leq 4} Q^I Q^J&&=\frac{1}{(n_+n_-)^2}\sum_{1\leq I<J\leq 4}p^I p^J&&\equiv\frac{1}{(n_+n_-)^2}\hat{P}^{(2)}\, ,\nonumber\\
\hat{Q}^{(3)}&\equiv \sum_{I=1}^{4} \prod_{J\neq I} Q^{J}&&=\frac{1}{(n_+ n_-)^3}\sum_{I=1}^{4}\prod_{J\neq I}p^J&&\equiv\frac{1}{(n_+ n_-)^3}\hat{P}^{(3)}\, ,\nonumber\\
 \hat{Q}^{(4)}&\equiv\prod_{I=1}^{4}Q^{I}&& =\frac{1}{(n_+ n_-)^4}\prod_{I=1}^{4}p^{I}&&\equiv\frac{1}{(n_+ n_-)^4}\hat{P}^{(4)}\, .
\end{alignat}
The $\hat{P}^{(a)}$ combinations are the ones which involve the integer magnetic charges and which we are ultimately interested in expressing everything in terms of, however in the intermediate computations the $\hat{Q}^{(a)}$ will be most useful.

{\bf Quartic invariant intermezzo} \\
The combinations $\hat{P}^{(a)}$ are the natural combinations arising from the quartic invariant for the STU model when considering a purely electric gauging with only magnetic charges. Let us recap the essential definitions of the quartic invariant to explain the connection. Let us define the charge vector and gauging parameters in the usual way
\be
\Gamma=\{ p^{\Lambda};q_{\Lambda}\}\, ,\quad G=\{g^{\Lambda};g_{\Lambda}\}\, ,
\ee
where the indices $\Lambda\in\{1,2,3,4\}=\{i, 4\}$ for consistency with our earlier notation. The quartic invariant is 
\begin{align}
I_4(\Gamma)&\equiv \frac{1}{4!}t^{ABCD}\Gamma_{A}\Gamma_{B}\Gamma_{C}\Gamma_{D}\nonumber\\
&=-(p^4 q_4-p^i q_i)^2 +4 q_1q_2 q_3 q_4+4 p^1p^2p^3p^4+4(p^1 p^2 q_1 q_2+p^1 p^3 q_1 q_3+p^2 p^3 q_2 q_3)\, ,
\end{align}
and we may obtain the symmetric tensor $t^{ABCD}$ via
\be
t^{ABCD}=\frac{\partial^4 I_4(\Gamma)}{\partial\Gamma_A\partial\Gamma_B\partial\Gamma_C\partial\Gamma_D}\, .
\ee
Using the tensor $t^{ABCD}$ we may extend $I_4$ to act on four distinct symplectic vectors as (note that the normalisation for $I_4(\Gamma)$ is different)
\be
I_4(W,X,Y,Z)=t^{ABCD} W_A X_B Y_C Z_D\, .
\ee

For a purely electric gauging, $g^{\Lambda}=0$ and for consistency with our normalisation of the gauge coupling which we set to 1, we take $g_\Lambda=1$, we find 
\begin{align}
\hat{P}^{(1)}&=\frac{1}{4!} I_4(\Gamma,G,G,G)\, ,
&&\hat{P}^{(2)}=\frac{1}{4^2}I_4(\Gamma,\Gamma,G,G)\, ,\nonumber\\
\hat{P}^{(3)}&=\frac{1}{4!}I_4 (\Gamma,\Gamma,\Gamma,G)\, ,
&&\hat{P}^{(4)}=\frac{1}{96}I_4 (\Gamma,\Gamma,\Gamma,\Gamma)=\frac{1}{4}I_4(\Gamma)\, .
\end{align}

\subsection{Expressing everything in terms of the quartic roots}

After this short intermezzo on the quartic invariant let us proceed with expressing these magnetic quantities we just defined in terms of the roots and orbifold weights. A tedious but otherwise simple computation shows that we may write these combinations of charges in terms of only the four roots, without the parameters $c^I$ appearing. With the assumptions that all the roots are both non-zero and not equal, in particular we do not assume any inequalities for the roots, these may be expressed as\footnote{These expressions also appear in the JHEP version of \cite{Couzens:2021rlk}.}
\begin{align}
\hat{Q}^{(4)}=&x^4\frac{w_1 w_4}{w_2 w_3}\, ,\label{eq:Qhat4}\\
\hat{Q}^{(3)}=&x^3\frac{w_1 w_4}{w_2 w_3}\bigg[(w_1+w_4)\frac{w_2 w_3}{w_1 w_4}+(w_1+w_4-2 w_2-2w_3)\bigg]\, ,\label{eq:Qhat3}\\
\hat{Q}^{(2)}=&x^2\frac{w_1 w_4}{w_2 w_3}\bigg[1+3(w_2+w_3)^2 -2 (w_2+w_3)(w_1+w_4)+w_1 w_4 \label{eq:Qhat2}\\
&-3 w_2 w_3
+\frac{1}{w_1 w_4}\Big(w_2 w_3+w_2 w_3 \sum_{1\leq I<J\leq 4}w_I w_J+(w_1+w_4-2w_2-2w_3)\sum_{I=1}^{4}\prod_{J\neq I} w_I\Big)\bigg]\, ,\nonumber\\
\hat{Q}^{(1)}=&x\Big[2(w_1+w_4)-(w_2+w_3)-\frac{w_1 w_4}{w_2w_3}(w_2+w_3)\Big]\, ,
\end{align}
where we defined
\be
x\equiv\frac{\Delta z(w_3-w_2)}{4\pi}>0\, .\label{eq:xdef}
\ee
First note that the free-energy, as given in \eqref{eq:freeE}, is proportional to `$x$'. Moreover at this point we could eliminate $\hat{Q}^{(1)}$ in favour of one of the orbifold weights $n_{\pm}$, however it useful to not do this until later. Finally, the Euler characteristic takes the form
\be
\chi=-\frac{x}{\sigma w_2 w_3}\frac{(w_3+\sigma w_2)(w_1w_4+\sigma w_2 w_3)-(1+\sigma)w_2 w_3 (w_1+w_4)}{\sigma w_2 w_3}\, .
\ee
A similar comment to that for $\hat{Q}^{(1)}$ applies here too.

\subsection{Period constraint}

Above we have managed to completely eliminate the parameters $c^{I}$ from the problem. We now want to solve the condition on the period \eqref{eq:period}. This will immediately imply that the expressions for $\hat{Q}^{(1)}$ and $\chi$ above take the canonical form in terms of $n_{\pm}$ defined previously. After some trivial substitutions the condition reduces to
\be
\sigma n_+ w_2 (w_3-w_1)(w_4-w_3)=n_- w_3(w_2-w_1)(w_4-w_2)\, ,
\ee
with both sides reassuringly positive. To solve this it is convenient to define
\be
w_1=\frac{1}{2} \Big(\alpha-\sqrt{\alpha^2 -4 \beta}\Big)\, ,~~ w_4=\frac{1}{2}\Big(\alpha+\sqrt{\alpha^2 -4 \beta}\Big)\, ,\quad \Leftrightarrow \quad w_1+w_4=\alpha\, ,~~ w_1 w_4=\beta\, ,
\ee
eliminating the two roots $w_1$ and $w_4$ in terms of $\alpha$ and $\beta$ everywhere and to solve the period constraint in terms of $\beta$. The solution is
\be
\beta=\frac{w_2 w_3 \big( n_-(w_2-\alpha)-\sigma n_+(w_3-\alpha)\big)}{n_+ w_2\sigma -n_- w_3}\, .
\ee
As a consistency check substituting this into the Euler characteristic and linear sum of the charges given above in terms of the roots gives the correct expressions for the twist and anti-twist:
\be
\hat{Q}^{(1)}=\sum_{I=1}^4 Q^I= \frac{1}{n_+}+\frac{\sigma}{n_-}\, ,\qquad \chi= \frac{1}{n_+}+\frac{\sigma^2}{n_-}\, ,
\ee 
where one should use that $\sigma^2=1$.

\subsection{Roots in terms of the magnetic charges}

Having eliminated $\beta$ in terms of the other roots and the orbifold weights, and thereby satisfied the period constraint, we may now eliminate $\alpha$. We do this by changing variables in favour of the variable $x$ defined above in equation \eqref{eq:xdef} rather than $\alpha$. The solution is
\be
\alpha=\frac{n_- w_3 (1+n_+ x w_3)-n_+ w_2 (\sigma +n_- x w_2)}{n_+n_- x(w_3-w_2)}\, ,
\ee
which we can now insert into the previous expressions to eliminate $\alpha$. 
So far we have eliminated the two roots $w_1$ and $w_4$ in favour of $n_\pm$; it remains to eliminate $w_2$ and $w_3$. It is once again useful to define new variables\footnote{Despite the simplicity of the coordinate change it is surprisingly powerful. One sees that it decouples the system of equations we are about to solve but for convenience suppress. One sees that $\gamma$ only appears in a single condition and not in all three.}
\be
w_2=\frac{\gamma-\delta}{2}\, ,\quad w_3=\frac{\gamma+\delta}{2}\, .
\ee
We now want to invert the expressions for the symmetric magnetic charge combinations, $\hat{P}^{(2)}\, ,\,  \hat{P}^{(3)} $ and $\hat{P}^{(4)}$ defined in \eqref{eq:Qhat4}-\eqref{eq:Qhat2} for the three variables $\{x,\gamma,\delta\}$ that we just introduced. It is a simple computation to insert this into mathematica and solve, we suppress the ugly intermediate results and just present the final result. We find four solutions differing by various signs. We may eliminate two out of four of the possibilities by noting that without loss of generality we have imposed $x>0$ and $\delta>0$.\footnote{They follow since we took $w_3>w_2$. We should also impose that $\gamma+\delta>0$ to ensure that $w_3>0$ as required, we will explain this further in the following section.} We are left with two distinct families of solutions parametrised by the sign $\tau=\pm 1$:
\begin{align}
x&=\frac{1}{\sqrt{2}n_+ n_-}\sqrt{\hat{P}^{(2)}-\sigma n_+ n_--\tau \sqrt{\big(\hat{P}^{(2)}-\sigma n_+ n_-\big)^2 -4 \hat{P}^{(4)}}}\, ,\label{eq:xsol}\\
\delta&= \tau \frac{x n_+ n_-(n_--\sigma n_+)}{\sqrt{\big(\hat{P}^{(2)}-\sigma n_+ n_-\big)^2 -4 \hat{P}^{(4)}}}\, ,\label{eq:deltasol}\\
\gamma &=\frac{ x n_+ n_- \big(2 \hat{P}^{(3)} + n_+ n_- (n_++\sigma n_-) -\sigma(n_++\sigma n_-)\hat{P}^{(2)}\big)}{\big(\hat{P}^{(2)}-\sigma n_+ n_-\big)^2 -4 \hat{P}^{(4)}}\, .\label{eq:gammasol}
\end{align}
Recall that the magnetic charges $p^I$, from which we construct the $P^{(I)}$'s, satisfy
\be
\sum_{I=1}^{4}p^I=n_-+\sigma n_+\, .\label{eq:psum}
\ee
The sign $\tau$ appears in two places; in the pre-factor of the inner square root in $x$ and the overall sign in  $\delta$. We could now use these expressions to determine the four roots in terms of the magnetic charges and orbifold weights, however, since we will firstly not need the roots any longer and given that the expressions are unwieldy we will refrain from presenting them here.\footnote{The reader may request a mathematica file in which the roots are written if they are curious.}
Given the above expressions we can immediately read off the free-energy in terms of the charges using \eqref{eq:freeE} and the solution for $x$ above. Note that there are two distinct classes for the form of the entropy depending on the sign of $\tau$. We will  show in the following sections that $\tau=-\sigma$.

\subsection{Regularity conditions}\label{sec:reg}

We have now managed to eliminate the roots in favour of the magnetic charges and orbifold weights. Before proceeding we must make sure that these new expressions are consistent with the ordering of the roots we imposed. Moreover, we still need to impose that the scalars are well-defined and strictly positive. In this final section will obtain the necessary conditions for the geometries to make well-defined, reducing these regularity conditions to a minimal set in terms of the physical parameters of the solution. 

First, we must require that the roots are ordered correctly, $w_1<w_2<w_3<w_4$ and $w_3>0$. For the twist we must also impose $w_2>0$ whilst for the anti-twist we must impose $w_2<0$. This naturally breaks the analysis into two distinct cases, in terms of $\gamma$ and $\delta$ these read 
\be
\begin{cases}\label{eq:casegds}
 \delta >|\gamma|>0\, ,\quad &\sigma=-1\\
\gamma>\delta>0\, ,\quad & \sigma=1
\end{cases}
\ee
which should be supplemented with $x>0, n_+>0$ and $n_->0$.

We also need to impose that the scalars are positive, which is equivalent to either taking $w_2-c^I>0$ or $w_3-c^I<0$ for all $c^I$. Since we have eliminated the $c^I$ completely this looks somewhat difficult to impose, however after some clever rewriting this is not the case. Using our favourite symmetric combinations it follows that
\be 
w_2-c^I>0\, ,\,\,\, \forall I \quad \Longleftrightarrow \quad 0< \begin{cases}
\sum_{I=1}^{4}(w_2-c^I)\, ,\\
\sum_{1\leq I<J\leq 4} (w_2-c^I)(w_2-c^J)\, ,\\
\sum_{I=1}^{4}\prod_{J\neq I}(w_2-c^J)\, ,\\
 \prod_{I=1}^{4}(w_2-c^I)\, ,
 \end{cases}
\ee
and similar conditions for $w_3-c^I<0$ are given by
\be 
w_3-c^I<0\, ,\,\,\, \forall I \quad \Longleftrightarrow \quad 0< \begin{cases}
-\sum_{I=1}^{4}(w_3-c^I)\, ,\\
\sum_{1\leq I<J\leq 4} (w_3-c^I)(w_3-c^J)\, ,\\
-\sum_{I=1}^{4}\prod_{J\neq I}(w_3-c^J)\, ,\\
 \prod_{I=1}^{4}(w_3-c^I)\, ,
 \end{cases}
\ee

The latter can all be expressed in terms of the roots of $f(w)$ by noticing that these combinations are precisely the ones that appear in derivatives of $P(w)$ evaluated at the roots $w_2$ and $w_3$. The positive scalar condition can then be split into the two sets:

{\bf First kind}
\begin{align}
&3w_2-w_1-w_3-w_4>0\, ,\quad \\
&1+(w_3-w_2)(w_4-w_2)+(w_2-w_1)(2 w_2-w_3-w_4)>0\, ,\nonumber\\
&2 w_2+(w_2-w_1)(w_2-w_3)(w_2-w_4)>0\, , \nonumber\\
& w_2^2>0\, ,\nonumber
\end{align}

{\bf Second kind}
\begin{align}
&w_1+w_2+w_4-3w_3>0\, ,\quad \\
&1+(w_3-w_2)(w_3-w_4)+(w_3-w_1)(2 w_3-w_2-w_4)>0\, ,\nonumber\\
&2 w_3+(w_3-w_1)(w_3-w_2)(w_3-w_4)<0\, , \nonumber\\
& w_3^2>0\, .\nonumber
\end{align}
Clearly in both cases the final condition is trivial. It remains to interpret these conditions in terms of the physical parameters.

Interestingly the first kind of positivity constraints for the scalars iimposes $0<\delta<3^{-1/2}$ in both cases and that $\gamma$ should always be positive. The refined conditions for the twist solutions, $\sigma=1$ give a single set of (non-trivial) bounds in terms of the auxiliary parameters
\begin{align}
&\sigma=1\, ,\quad 0<\delta<\text{min}(3^{-1/2},\gamma)\, ,\quad n_+> \frac{\gamma+\delta}{2x(1-3 \delta^2)}\, ,\nonumber\\
&\frac{2 n_+ (\gamma-\delta)}{\gamma+\delta+2 n_+ x(1+\delta^2)}<n_-<\frac{n_+ (\gamma-\delta)}{\gamma+\delta+4 n_+ x \delta^2}\, .\label{eq:twistrangegen}
\end{align}
One can show that $n_-<n_+$ for a well defined solution. This asymmetry is an artefact of fixing $w_3>w_2>0$, and also means that for the second kind of positive scalars there are no solutions for the twist solutions. We may interchange the order of the orbifold parameters by using the symmetry in \eqref{eq:symmetry}, which interchanges $n_{\pm}$. We will further refine these conditions in section \ref{sec:twist} in terms of the magnetic charges.

The anti-twist bounds ($\sigma=-1)$ are more involved and must be broken into two cases. Case 1 is
\begin{align}
&0<\delta<\frac{1}{\sqrt{3}}\, ,\quad \delta^3<\gamma<\delta-2 \delta^3\, ,\quad n_+> -\frac{n_-(\gamma+\delta)}{\gamma+\delta(4 n_- \delta x-1)}\, ,\quad \nonumber\\
&\frac{\delta-\gamma}{x(1+\delta^2)}<n_-<\frac{1}{4x \delta^3}\text{min}\big(2 \delta^3, \delta-\gamma\big)\, ,
\end{align}
and case 2 is
\begin{align}
&0<\delta<\frac{1}{\sqrt{3}}\, ,\quad \delta^3<\gamma<\delta\, ,\quad \frac{\gamma-\delta}{2x(\delta^2-1)}<n_-<\frac{1}{2x (1+\delta^2)}\text{min}\big(2(\delta-\gamma),\delta(1+\delta^2)\big)\, ,\nonumber\\
&\frac{n_- (\gamma+\delta)}{\delta (1-4 n_- x\delta )-\gamma}<n_+<\frac{n_- (\gamma+\delta)}{2(\delta- \gamma-n_- x(1+\delta^2))} \, .
\end{align}

For anti-twist solutions there are a large number of possibilities, the most notable are the two simple ranges:
\begin{align}
&\sigma=-1\, ,\quad\delta> \gamma>0\, ,\quad  \delta>1\, ,\quad x>0\, ,\quad n_+>0\, ,\quad 2 x n_-> \delta\, ,\nonumber\\
&\sigma=-1\, ,\quad \delta>\sqrt{3}\, ,\quad 0>\gamma>-\delta\, ,\quad n_+>0\, ,\quad 2 n_- x>\delta\, ,
\end{align}

We are now in a position to study the parameter space of admissible solutions. In the following we will obtain bounds on the free parameters of the solution in terms of the symmetric charge combinations $\hat{P}^{(I)}$. In principle one can obtain conditions on the magnetic charges $p^I$ rather than these symmetric combinations, it is certainly possible in mathematica, however the resultant expressions are far more complicated than the simple(ish) bounds that we find here. 
As we will see shortly we do not encounter any issues in finding explicit charge configurations, $\{p^I, n_{\pm}\}$, using the bounds in terms of the symmetric combinations so we are not losing anything by not expressing them in terms of the $p^I$. We will first consider the parameter space of anti-twist solutions in section \ref{sec:antitwist} before moving on to the twist solutions in section \ref{sec:twist}.

%%%%%%%%%%%%%%%%%%%%%%%%%%%%%%%%%%%%%%%%%%%
%%%        Anti-twist
%%%%%%%%%%%%%%%%%%%%%%%%%%%%%%%%%%%%%%%%%%%

\section{Anti-twist solutions}\label{sec:antitwist}

We now want to study the parameter space of the solutions in terms of the magnetic charges for the anti-twist solution. We will first study the anti-twist solution, $\sigma=-1$ in this section before moving on to the twist solution in the following section.
Solutions of this type have previously been studied in the $X^0X^1$ truncation in \cite{Couzens:2021rlk,Ferrero:2021ovq} and the Einstein--Maxwell truncation in \cite{Ferrero:2020twa}, the latter two references also allow for rotation. Here we extend the analysis to the 4-magnetic charge case. It would be interesting to extend this to study the 4-charge rotating solution which is currently unknown.

Recall that we require $\delta>0$ and therefore it follows that only the solution with $\tau=1$ is valid, the other leads to $w_2>w_3$ which we must avoid. In addition to imposing that the roots are real which requires 
\be
\Big(n_+ n_-+\hat{P}^{(2)}\Big)^2 > 4 \hat{P}^{(4)}\, ,\qquad \hat{P}^{(2)}+n_+ n_->0\, ,
\ee
we must also impose the constraints in \eqref{eq:casegds} and also those imposing the positive scalars. In the following we will present only the conditions for the roots to have the correct form, and instead impose the positive scalar conditions for each of the charge configurations whilst performing the search:\footnote{The final constraint is actually implied by the first two however since it is not immediately obvious that this is true we present it for ease of understanding.}
\begin{align}
&0<\hat{P}^{(4)}<\frac{\Big(n_-\big(n_+ n_-+\hat{P}^{(2)}\big)-\hat{P}^{(3)}\Big)\Big(\hat{P}^{(3)}+n_+\big( n_+ n_- +\hat{P}^{(2)}\big)\Big)}{(n_++n_-)^2}\, ,\\
&-n_+\Big(n_+ n_-+\hat{P}^{(2)}\Big)<\hat{P}^{(3)}<n_-\Big(n_+ n_-+\hat{P}^{(2)}\Big)\, ,\\
&0<n_+ n_-+\hat{P}^{(2)}\, .
\end{align}
Note that for
\be
\hat{P}^{(3)}=\frac{1}{2}(n_--n_+)\Big(n_+ n_-+\hat{P}^{(2)}\Big)\, ,
\ee
we have $\gamma=0$ and the two end-point roots satisfy $w_2=-w_3$.

It is clear from the constraints that we require the same number of positive as negative roots since $\hat{P}^{(4)}$ is positive definite. The constraints on having well-defined roots seem can be solved relatively easily for any combination of an even number of positive magnetic charges. The positivity constraints on the scalars are far more restrictive and seem to indicate that we must take only negative magnetic charges if we impose $w_2-c^I>0$ and only positive charges if we take $w_3-c^I<0$.

\subsection{Class 1: $n_+>n_->0$}

\begin{figure}[h!]
\centering
  \includegraphics[width=0.5\linewidth]{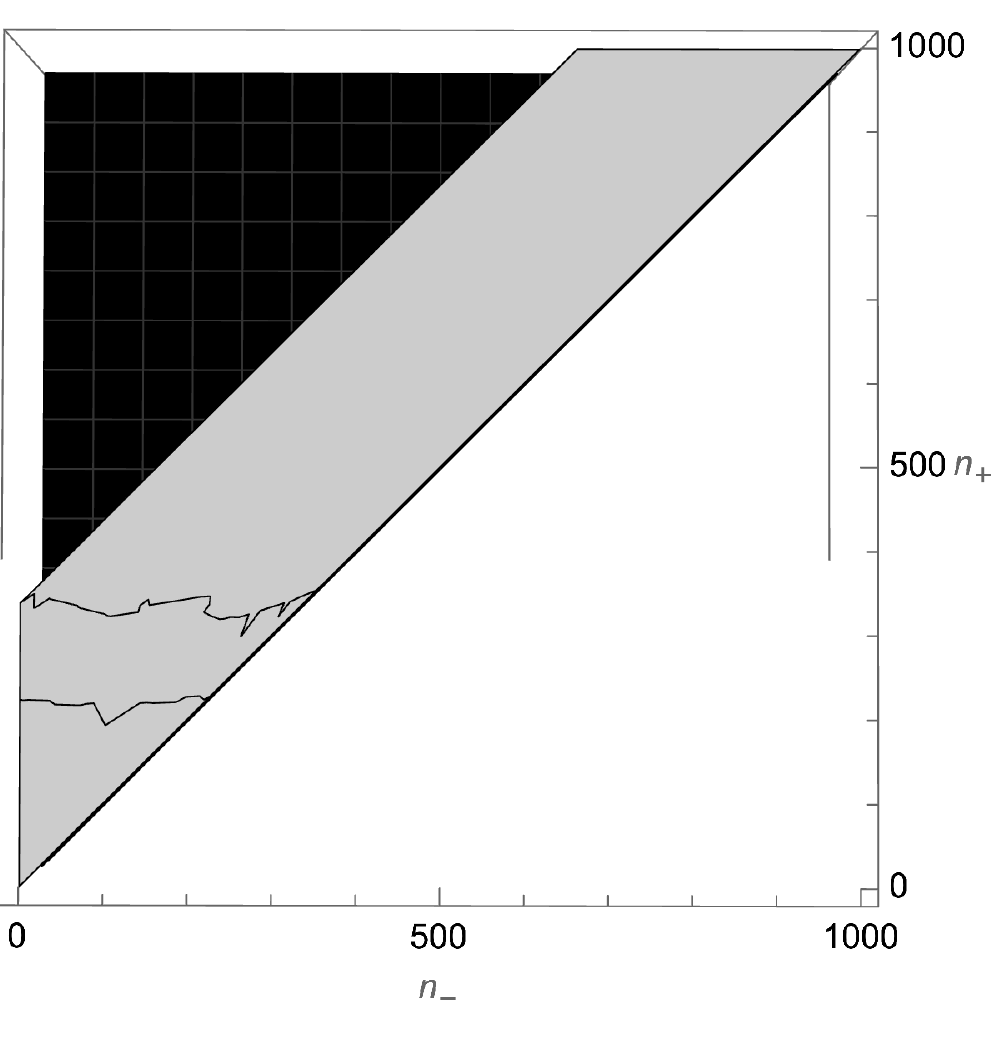}
  \captionsetup{width=.95\linewidth}
  \caption{\textit{We plot the domain of validity of the anti-twist solutions for the magnetic charges $\{ p_2,p_3,p_4\}=\{-11, -17,-319\}$. The black region satisfies all the necessary inequalities for a well-defined anti-twist solution with $w_2-c^I>0$. The grey and white regions give inadmissible solutions. We find similar plots after taking all positive magnetic charges with $w_3-c^I<0$ and with the roles of $n_+$ and $n_-$ interchanged.}}
  \label{fig:antitwistplot}
\end{figure}

Let us consider the first case, picking three seed magnetic charges, all negative. We end up with infinite families of solutions, see figure \ref{fig:antitwistplot} . Examples of triplets of seed magnetic charges, the fourth is fixed by satisfying \eqref{eq:psum}, and the bounds on $n_{\pm}$ are
\begin{align}
\{p_2,p_3,p_4\}=
\begin{cases}
\{-11,-17,-319\}\, ,\quad &n_+>347+n_-\, ,\\
\{-p^2,-p^2,-p^2\}\, ,\quad & n_+>3 p^2+n_-\, ,\\
\end{cases}
\end{align}
More generally for the charge configuration
\be
\{-m_1^2,-m_2^2,-m_3^2\}\, ,
\ee
the necessary bound in almost all of parameter space is
\be
n_+> m_1^2+m_2^2+m_3^2+n_-\, ,
\ee
though we could not prove this in general. Despite not being able to prove the form of the lower bound for large enough $n_{+}>n_{-}$ for the large number of data points we test we always found a solution.

\subsection{Class 2: $n_->n_+>0$}

For the second way of enforcing positive scalars we have the opposite scenario. For all positive magnetic charges we have infinite families of solutions with $n_->n_+$. In fact, we may obtain consistent solutions by flipping the sign of the negative seed charges above whilst also flipping the role of $n_+$ and $n_-$, this is precisely the flip symmetry \eqref{eq:symmetry}. Consistent charge configurations are then
\begin{align}
\{p_2,p_3,p_4\}=
\begin{cases}
\{11,17,319\}\, ,\quad &n_->347+n_+\, ,\\
\{p^2,p^2,p^2\}\, ,\quad & n_->3 p^2+n_+\, ,\\
\end{cases}
\end{align}
and for the general charge configuration
\be
\{m_1^2,m_2^2,m_3^2\}\, ,
\ee
the necessary bound in almost all of $(n_-,n_+)$ parameter space is
\be
n_-> m_1^2+m_2^2+m_3^2+n_+\, .
\ee
As before despite not being able to prove the form of the lower bound, for all positive seed magnetic charges and for large enough $n_{-}>n_{+}$ we find a solution having tested this on a large number of data points.

We have shown that there is a plethora of anti-twist solutions, for both $n_+>n_-$ and $n_->n_+$. All of the solutions we have found have involved either four positive magnetic charges or four negative magnetic charges, we have not been able to rule out conclusively an even mixture of both positive and negative magnetic charges in the general 4-charge solution however it is possible to rule this possibility out for the three consistent truncations. Numerics seems to support that no such solutions exist either for the unrestricted multi-charge case but it would be interesting to prove this conclusively.

Before we move on to the twist solutions let us provide the closed form expression for the free-energy of the four-charge solution,\footnote{The same result appears in the JHEP version of \cite{Couzens:2021rlk}.}
\be
\mathcal{F}=\frac{2 \pi}{3}N^{3/2}\frac{\sqrt{n_+ n_- +\hat{P}^{(2)}-\sqrt{\big(n_+ n_-+\hat{P}^{(2)}\big)^2 -4 \hat{P}^{(4)}}}}{n_+ n_-}\, .
\ee
We see that in both the $X^0X^1$ and Einstein--Maxwell truncation this reduces correctly to the free-energy given in \cite{Ferrero:2021ovq}. We could insert the expressions for the $\hat{P}^{(I)}$ in terms of the quartic invariant at this point however we will refrain from doing this. 
%%~~~~~~~~~~~~~~~~~~~~~~~~~~~~~~~~~~~~~~~~~~~~~~~~~~~~~~~~~~~~~~~~~~~~

%%~~~~~~~~~~~~~~~~~~~~~~~~~~~~~~~~~~~~~~~~~~~~~~~~~~~~~~~~~~~~~~~~~~~~

%%%%%%%%%%%%%%%%%%%%%%%%%%%%%%%%%%%%%%%%%%%
%%%        Twist
%%%%%%%%%%%%%%%%%%%%%%%%%%%%%%%%%%%%%%%%%%%

\section{Twist solutions}\label{sec:twist}

Having studied the anti-twist solutions in the previous section let us turn our attention to the twist solutions. Recall from section \ref{sec:reg} that we must take $n_+>n_->0$ and therefore in order for $\delta>0$ we must fix $\tau=-1=-\sigma$. We reiterate that the apparent asymmetry is due to our choice $w_3>w_2>0$ and the other option may be obtained by using the flip symmetry \eqref{eq:symmetry} we will focus on the case with $w_3>0$ and therefore $n_+>n_-$.

\begin{figure}[h!]
\centering
  \includegraphics[width=0.5\linewidth]{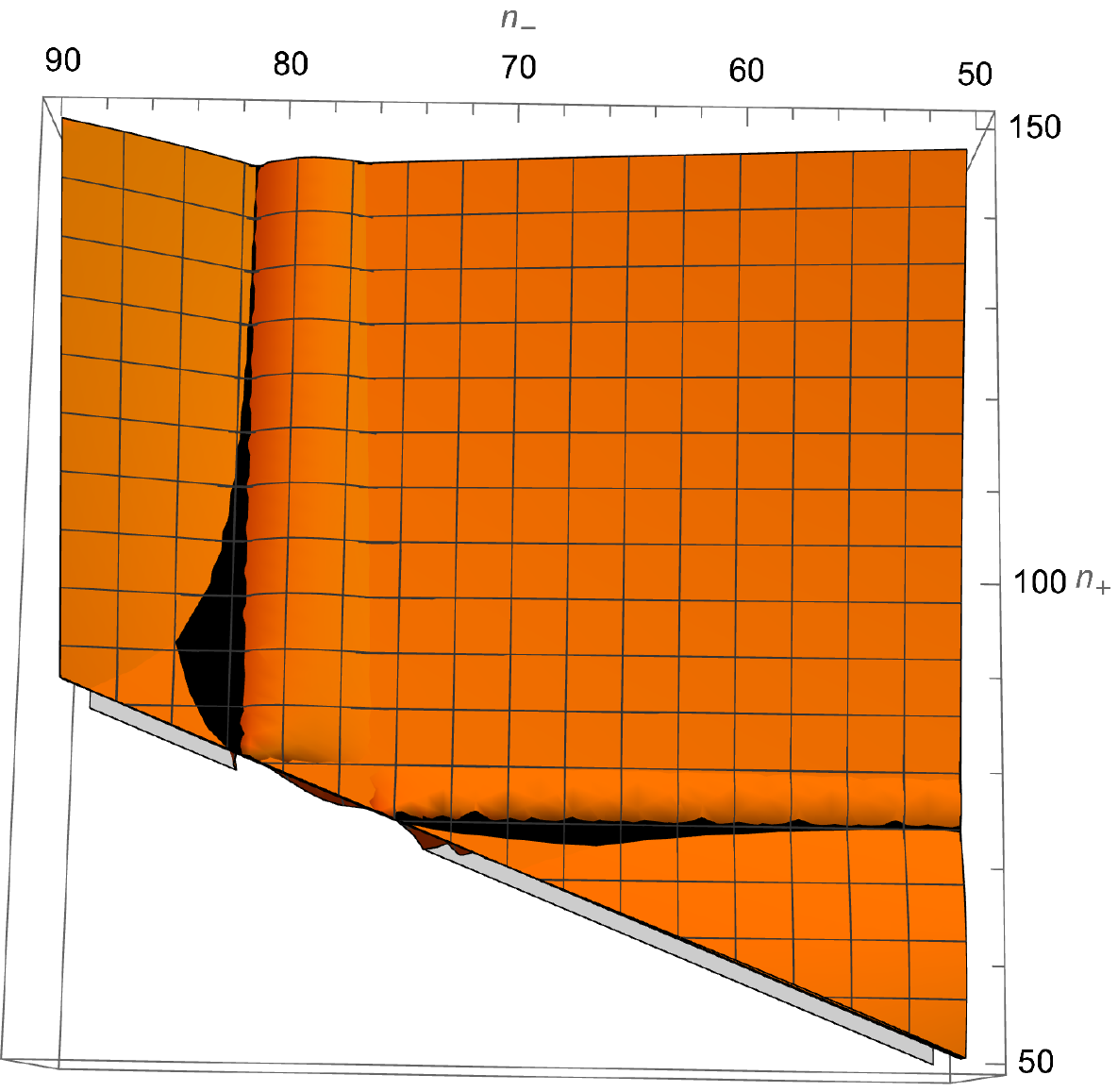}
  \captionsetup{width=.95\linewidth}
  \caption{\textit{We plot the domain of validity for the magnetic charges $\{ p_1,p_2,p_3,p_4\}=\{-87+n_++n_-,5,11,71\}$ for regime 1 of class 2 twist solutions. The plot black islands are areas in $(n_+,n_-)$ parameter space where the order of the roots has been enforced. Imposing in addition the positivity of the scalars lifts these regions leaving just orange regions and no solution. This type of behaviour is common to both regime 1 and regime 2 and therefore we shall ignore them in the following.  }}
  \label{fig:twistcase2reg1plot}
\end{figure}

The hope of finding twist solutions with $w_3>0$ therefore rest on finding solutions with $n_+>n_->0$.\footnote{The case A numerical solution in \cite{Ferrero:2021etw} is indeed in this class with $n_+\sim 3.4 n_-$. } The constraints for the roots to have the correct form allows for three regimes which may be found below. Imposing on top that the scalars are positive definite leads to a far more constrained system, which is also more difficult to write down due to a larger number of possibilities and redundancies. However, by first studying the solutions for just the correct roots we find that for two of the three regimes only small islands of solutions can be found, see for example figure \ref{fig:twistcase2reg1plot}. Imposing on top of this the scalar positivity constraints leads to these islands vanishing for all charge configurations that we checked. We presume that in these two regimes there are no solutions but we have been able to rule this out completely. The third regime is far kinder to us and leads to infinite families of solutions, reminiscent of how simple it was to find the anti-twist solutions.  
The three regimes for solutions with the correct root structure are given below. It is the third regime where we can find infinite families of solutions and therefore we will focus on that case. 

{\bf Regime 1}
The first regime has a strictly positive $\hat{P}^{(4)}$ subject to the inequalities
\begin{align}
&\hat{P}^{(2)}>n_+ n_-\, ,\quad \frac{1}{2}(n_++n_-)(\hat{P}^{(2)}-n_+ n_-)<\hat{P}^{(3)}<n_+(\hat{P}^{(2)}-n_+ n_-)\, ,\quad\\
&  -\frac{\big(\hat{P}^{(3)}-n_-(\hat{P}^{(2)}-n_+ n_-)\big)\big(\hat{P}^{(3)}-n_+(\hat{P}^{(2)}- n_+ n_-)\big)}{(n_+-n_-)^2}<\hat{P}^{(4)}<\frac{1}{4} (\hat{P}^{(2)}-n_+ n_-)^2\, . \nonumber
\end{align}

{\bf Regime 2} The second regime allows for both a positive and negative $\hat{P}^{(4)}$ subject to
\begin{align}
&\hat{P}^{(2)}>n_+ n_-\, ,\quad n_+(\hat{P}^{(2)}-n_+ n_-)<\hat{P}^{(3)}\, ,\quad\\
& -\frac{\big(\hat{P}^{(3)}-n_-(\hat{P}^{(2)}-n_+ n_-)\big)\big(\hat{P}^{(3)}-n_+(\hat{P}^{(2)}- n_+ n_-)\big)}{(n_+-n_-)^2}<\hat{P}^{(4)}<\frac{1}{4} (\hat{P}^{(2)}-n_+ n_-)^2\, . \nonumber
\end{align}

{\bf Regime 3} The third and final regime has a strictly negative $\hat{P}^{(4)}$ satisfying
\begin{align}
&\hat{P}^{(2)}<n_+ n_-\, ,\quad \hat{P}^{(3)}> n_-(\hat{P}^{(2)}-n_+ n_-)\, ,\\
& -\frac{\big( \hat{P}^{(3)}-n_-(\hat{P}^{(2)}-n_+ n_-)\big)\big(\hat{P}^{(3)}- n_+(\hat{P}^{(2)}-n_+ n_-)\big)}{(n_+-n_-)^2}<\hat{P}^{(4)}<0\, .\nonumber
\end{align}

\subsection{Infinite families of solutions}

This final regime requires an odd number of positive charges since $\hat{P}^{(4)}<0$. For the anti-twist solutions we had $\hat{P}^{(4)}>0$ in the bountiful region. We are once again able to find infinite numbers of solutions for generic seed magnetic charges which are all negative. Amazingly we find that for any seed solution with all three magnetic charges negative any choice of $n_+>n_->0$ gives a valid solution. Some of the explicit configurations that we have checked are 
\begin{align}
\{p_2,p_3,p_4\}=\begin{cases}
\{- 1, -2,-3\}\, ,\qquad &n_+>n_->0\, ,\\
\{- 132, -589,-3554\}\, ,\qquad &n_+>n_->0\, ,\\
\{- 21, -1993,-1345245\}\, ,\qquad &n_+>n_->0\, ,\\
\{-p^2,-p^2,-p^2\}\, , \quad &n_+>n_->0
\end{cases}
\end{align}
though there are plenty more with higher and higher orders. The important constraint is that all three seed charges are negative and it follows that we end up with three negative and one positive magnetic charge. Though we did not algebraically prove this the large number of data points we have tested indicates that this should true and we conclude that infinite families of solutions exist for three negative and one positive magnetic charges for $n_+>n_->0$. 
Note that this analysis agrees with the fact that in neither Einstein--Maxwell nor the $X^0 X^1$ truncation one can find a twist solution as shown in \cite{Ferrero:2021etw,Couzens:2021rlk}. In these truncations the quartic polynomial $f(w)$ is simple enough that one can prove this fully algebraically.

Given the flip symmetry we can obtain solutions where $n_->n_+$ by taking three positive seed magnetic charges and one negative. We have checked this explicitly and the bounds agree with the transformed ones discussed above,

We finish this section by giving the closed form expression for the free-energy for twist solutions,
\be
\mathcal{F}=\frac{2 \pi}{3}N^{3/2}\frac{\sqrt{\hat{P}^{(2)}-n_+ n_- +\sqrt{\big(\hat{P}^{(2)}-n_+ n_-\big)^2 -4 \hat{P}^{(4)}}}}{n_+ n_-}\, .
\ee
We emphasise that this is distinct to the expression for the anti-twist free-energy by three signs.

\section{Conclusion}

We have studied the possibility of realising both the twist and anti-twist for the multi-charge AdS$_2\times\mathbb{WCP}^{1}_{[n_+,n_]}$ solutions. We have constructed infinite classes of both solutions, parametrised by the choice of seed magnetic charges, providing insurmountable evidence for their existence, in agreement with \cite{Ferrero:2021etw}. 
For twist solutions we find infinite families of solutions when three of the four magnetic charges are either all positive or all negative. Whilst for anti-twist solutions we find that all magnetic charges are either positive or all negative with the sign correlated to the magnitude of the orbifold weights. It would be interesting to prove that these are the only possibilities. A non-extensive numerical analysis seems to confirm this but we were unable to present an algebraic proof.

We have provided compact and explicit expressions for the free-energy of the solutions expressed in terms of the magnetic charges and orbifold weights. 
In \cite{Faedo:2021nub}, they conjectured a form for the off-shell free-energy wrapped brane spindle solutions in various dimensions. Their conjecture for the off-shell free energy for M2-branes is\footnote{We have rewritten the charges into our notation and swapped the definitions of $F$ and $\mathcal{F}$ in \cite{Faedo:2021nub} for consistency with our notation.}
\be
\mathcal{F}^{-\sigma}=-\frac{\sqrt{2}\pi}{3}N^{3/2} \frac{1}{\epsilon}\Big(F(\varphi_I+\epsilon Q_I)-\sigma F(\varphi_I-\epsilon Q_I)\Big)\,,
\ee
where 
\be
\sum_{I=1}^{4}\varphi_I-\frac{n_+-\sigma n_-}{n_+ n_-} \epsilon=2\, ,
\ee
and, like here the charges satisfy the constraint
\be
\sum_{I=1}^{4}Q_I=\frac{n_++\sigma n_-}{n_+ n_-}\, .
\ee
As explained in \cite{Faedo:2021nub} this should be extremised for $\varphi_I$ and $\epsilon$ subject to the constraint between $\varphi_I$ and $\epsilon$.  The parameter $\epsilon$ is a fugacity associated to the rotational symmetry of the spindle whilst the $\varphi_I$ are fugacities for the U$(1)^4$ symmetry. It is a feature of the spindle (and disc) geometries that the R-symmetry mixes with the rotational symmetries of the compactification surface and therefore requires the inclusion of the fugacity $\epsilon$ which would not otherwise appear for a static geometry like the ones we are considering here. 

Given our expressions we find that the free-energy of the multi-charge spindle solutions for twist and anti-twist solutions can be written as
\be
\mathcal{F}=\frac{2\pi}{3n_+ n_-}N^{3/2}\sqrt{\hat{P}^{(2)}-\sigma n_+ n_- +\sigma \sqrt{\big(\hat{P}^{(2)}-\sigma n_+ n_-\big)^2 - 4 \hat{P}^{(4)}}}\, .
\ee
It would be interesting to recover this result from extremising the above functional and to understand how this latter constraint arises.

%%%%%%%%%%%%%%%%%%%%%%%%%%%%%%%%%

\section*{Acknowledgments}

It is a pleasure to thank Hyojoong Kim, Nakwoo Kim, Yein Lee, Myungbo Shim, Minwoo Suh, Koen Stemerdink and Damian van de Heisteeg for useful discussions. I would also like to thank Pietro Ferrero, Jerome Gauntlett and James Sparks for comments on an earlier version of the draft and Andrea Boido for pointing out a typo. CC is supported by the National Research Foundation of Korea (NRF) grant 2019R1A2C2004880.

%%%%%%%%%%%%%%%%%%%%%%
\appendix

%%%%%%%%%%%%%%

\section{Killing spinors on the spindle}\label{app:spinors}

In this section we will study the Killing spinors of the four-dimensional solution. For the conventions of the supersymmetry transformations we take the general form of the Killing spinor equations of 4d $\mathcal{N}=2$ gauged supergravity in \cite{Lauria:2020rhc} with the prepotential
\be
F=-\ii \sqrt{X^{(1)}X^{(2)}X^{(3)}X^{(4)}}\, .
\ee
We work in the gauge where 
\be
X^{(1)}X^{(2)}X^{(3)}X^{(4)}=1\, ,
\ee
and define the physical scalars via
\be
X^{(I)}=\me^{\vec{v}^I\cdot \vec{\phi}}\, ,
\ee
with 
\be
\vec{v}^1=\tfrac{1}{2}\{1,-1,-1\}\, ,\quad \vec{v}^2=\tfrac{1}{2}\{-1,1,-1\}\, ,\quad \vec{v}^3=\tfrac{1}{2}\{-1,-1,1\}\, ,\quad \vec{v}^4=\tfrac{1}{2}\{1,1,1\}\, ,\quad 
\ee

The resultant gravitino Killing spinor equation is\footnote{The Killing spinor equations are given in terms of symplectic Majorana spinors, with $\epsilon^{i}$ of positive chirality and $\epsilon_i$ of negative chirality. It is convenient to rewrite the Killing spinor equations in terms of the Dirac spinor $\epsilon=\epsilon^1+\epsilon_2$, which we will take from now on. }
\be
\delta \Psi_{\mu}=\bigg[ \nabla_{\mu}-\frac{\ii}{4}\sum_{I=1}^{4} A^{I}_{\mu} +\frac{1}{8}\sum_{I=1}^{4} X^{(I)} \gamma_{\mu}+\frac{\ii}{8}\sum_{I=1}^{4} \big(X^{(I)})^{-1} \slashed{F}^{(I)}\gamma_{\mu}\bigg]\epsilon
\ee
whilst the three gaugino Killing spinor equations are
\be
\delta \lambda_i= \bigg[\slashed{\partial}\phi_i -\sum_{I=1}^{4} \vec{v}^{I}_{i} X^{(I)}-{\ii}\sum_{I=1}^{4}\vec{v}^{I}_{i} (X^{(I)})^{-1}\slashed{F}^{I}\bigg]\epsilon\, .
\ee
Working in components, on AdS$_2$ we find ($a=0,1$)
\begin{align}
\delta\Psi_{a}&= \bigg[\hat{\nabla}_{a}-\frac{\ii \alpha}{2}\gamma_{23}\gamma_{a}+\frac{P'(w)}{4 \sqrt{P(w)}}\Big(\frac{\alpha}{2}-\frac{\sqrt{f(w)}}{2\sqrt{P(w)}}\gamma_{2}+\frac{\ii \alpha w}{2\sqrt{P(w)}}\gamma_{23}\Big)\gamma_a\bigg]\epsilon\, ,
\end{align}
with 
$\hat{\nabla}_a$ the covariant derivative on unit radius AdS$_2$ and the curved indices with respect to the conformally rescaled metric which removes the overall conformal factor in \eqref{eq:NHmetric}. The parameter $\alpha$ is either $\pm1$ depending on whether $w-c^I>0, ~\alpha=1$ or $w-c^I<0, ~\alpha=-1$. This is related to the two different ways of enforcing positive scalars. 
The other two Killing spinor are
\begin{align}
\delta \Psi_w&=\bigg[\partial_w -\frac{\ii \alpha}{2}\gamma_{23}\gamma_w +\frac{\alpha P'(w)}{4\sqrt{P(w)}}\Big(\frac{1}{2}+\frac{\ii w}{2\sqrt{P(w)}}\gamma_{23}\Big)\gamma_w\bigg]\epsilon\, ,\\
\delta\Psi_z&=\bigg[\partial_z+\frac{f(w)P'(w)-2P(w)f'(w)}{8 P(w)^{3/2}}\gamma_{23}-\frac{\ii\alpha}{2}\gamma_{23}\gamma_z+\frac{\ii}{4}\Big(\frac{w P'(w)}{P(w)}-\sum_I n^I\Big)\nonumber\\
&\quad+\frac{\alpha P'(w)}{4\sqrt{P(w)}}\Big(\frac{1}{2}+\frac{\ii w}{2\sqrt{P(w)}}\gamma_{23}\Big)\gamma_z\bigg]\epsilon\, .
\end{align}
We have allowed for an arbitrary gauge choice $\delta A^I=n^I \dd z$ for the gauge fields parametrised by the $n^I$, see \cite{Ferrero:2021etw} for a detailed discussion of gauge choices.

Let us now solve these conditions. From the Killing spinor equation along AdS$_2$ we see that we must impose that the spinor satisfies the projection condition
\be
\Big(\frac{\alpha}{2}-\frac{\sqrt{f(w)}}{2\sqrt{P(w)}}\gamma_{2}+\frac{\ii \alpha w}{2\sqrt{P(w)}}\gamma_{23}\Big)\epsilon=0\, ,
\ee
which implies
\be
\Big[\hat{\nabla}_a-\frac{\ii \alpha}{2}\gamma_{a}\gamma_{23}\Big]\epsilon=0\, .
\ee
Let us take the gamma matrices
\be
\gamma_{0}=\ii \sigma_2\otimes \sigma_3\, ,\quad \gamma_1=\sigma_3\otimes \sigma_3\, ,\quad \gamma_2=1_{2\times2}\otimes \sigma_1\, ,\quad \gamma_3=1_{2\times2}\otimes\sigma_2\, ,
\ee
then $\gamma_a\gamma_{23}=\ii\,\rho_a\otimes1_{2\times2}$, with $\rho_a$ the 2d gamma matrices for AdS$_2$ and therefore the Killing spinor on AdS$_2$ reduces to 
\be
\Big[\hat{\nabla}_a+\frac{ \alpha}{2}\rho_a\otimes1_{2\times2}\Big]\epsilon=0\, .\label{eq:AdS2KSE}
\ee
We should decompose the 4d spinor in terms of AdS$_2$ Killing spinors. There are two inequivalent Killing spinor equations that we can construct depending on the sign of $\alpha$. 
In terms of these spinors we can decompose the 4d spinors as\footnote{The $\pm$ index of the $\theta$'s confers no information about a projection.}
\be
\epsilon_\pm =\eta_\pm \otimes\theta_\pm\, ,\qquad \text{with} \qquad \sigma_3\theta_{\pm}=\pm\theta_{\pm}\, ,
\ee
with 
$\eta_-$ solving \eqref{eq:AdS2KSE} for $\alpha=1$ and $\eta_+$ solving \eqref{eq:AdS2KSE} for $\alpha=-1$.
If we put the following metric on AdS$_2$
\be
\dd s^2(\text{AdS}_2)=-r^2\dd t^2+\frac{\dd r^2}{r^2}\, ,
\ee
then the Killing spinors on AdS$_2$ are
\be
\eta_+=\big(\sqrt{r}(c_1+c_2 t),r^{-1/2}c_2\big)\, ,\quad \eta_-=\big(c_1 r^{-1/2},\sqrt{r}(c_1 t+c_2)\big)\, .
\ee
We may now solve the Killing spinor equations by first solving the projection condition, and then solving for the remaining component.

Let us first consider $\alpha=1$. We expect to construct a spinor utilising the $\eta_-$ spinor on AdS$_2$. We find the solution 
\be
\theta_-=P(w)^{-1/8}\me^{\tfrac{\ii z}{4}\big(2-\sum_I n^I\big)}\bigg(\sqrt{\sqrt{P(w)}+w}\, ,~-\sqrt{\sqrt{P(w)}-w}\,\bigg)\, .
\ee
For $\alpha=-1$ we find that the AdS$_2$ spinor we must take is $\eta_+$ and the spinor $\theta_+$ is 
\be
\theta_+=P(w)^{-1/8}\me^{\tfrac{\ii z}{4}\big(2-\sum_I n^I\big)}\bigg(\sqrt{\sqrt{P(w)}+w}\, ,~\sqrt{\sqrt{P(w)}-w}\,\bigg)=\sigma_3\cdot \theta_-\, .
\ee
Note that these spinors agree with the ones found in \cite{Ferrero:2021etw} once the simple redefinitions mapping between the two solutions are imposed. 

Let us now consider the differences between the two types of twist. Since $\theta_+=\sigma_3 \theta_-$ we can restrict to considering only $\theta_+$ without loss of generality. First note that the product of the entries of the spinor is precisely the function $\sqrt{f(w)}$ that is
\be
\sqrt{f(w)}=\sqrt{\sqrt{P(w)}+w}\, \cdot\sqrt{\sqrt{P(w)}-w}\, .
\ee
Recall that at a root, $w_*$ of $f(w)$ we have
\be
P(w_*)=w_*^2\, .
\ee
Therefore we see that at one of the poles of the spindle one entry of the spinor vanishes, but not both. Let us reinstate the two roots, $w_2=\sigma |w_2|$ and $w_3>0$. Then at $w_3$ we have
\be
\theta_+(w_3)=P(w_3)^{-1/8}\me^{\tfrac{\ii z}{4}\big(2-\sum_I n^I\big)}\sqrt{2}\big(\sqrt{w_3}\, ,~0\big)\, ,
\ee
whilst at $w_2$ we have
\be
\theta_+(w_2)=P(w_2)^{-1/8}\me^{\tfrac{\ii z}{4}\big(2-\sum_I n^I\big)}\sqrt{2}\begin{cases}
\big(\sqrt{w_2}\, ,~0\big)\, ,&\text{twist} ~~(\sigma=1)\\
\big(0\, ,~\sqrt{-w_2}\big)\, ,&\text{anti-twist} ~~(\sigma=-1)
\end{cases}\, .
\ee
We therefore see that the twist and anti-twist solutions have Killing spinors with different properties.\footnote{Note that the disc preserves supersymmetry in an altogether different way since at the boundary of the disc, located at $w=0$, the Killing spinor vanishes as $w^{1/8}$.}

\section{Full black hole solution}\label{app:fullblackhole}

We present a full black hole solution with near-horizon given by the solution studied in the main text. Having presented the full black hole solution we take the near-horizon limit and with a change of coordinates show that it recovers the near-horizon solution studied in the main text. We will not check explicitly the supersymmetry of the full black hole solution, instead we will require that it is extremal and has the supersymmetric solution studied in the main text as near-horizon geometry. Of course this is necessary but not sufficient for the full black hole solution to preserve supersymmetry however we will content ourselves with this in this work. 

The solution was originally found in \cite{Lu:2014sza} and was conjectured to give rise to the near-horizon solutions we study in this work in \cite{Ferrero:2021ovq}. Dualising the solution \cite{Lu:2014sza} to have only magnetic charges we have\footnote{We have changed the definitions of some of the functions and parameters to remove some of the redundancy in the definitions and potential confusion with previous expressions. }
\begin{align}
\dd s^2_4&=\frac{\sqrt{\mathcal{F}(x) \mathcal{H}(y)}}{\alpha^2(y-x)^2}\bigg[ -Y(y)\dd t^2+\frac{\mathcal{H}(y)}{Y(y)}\dd y^2+\frac{\dd x^2}{X(x)}+\frac{X(x)}{\mathcal{F}(x)}\dd\phi^2\bigg]\, ,\label{eq:fullBH}\\
X^{(I)}&=\bigg[\frac{h_I(y)^4}{\mathcal{H}(y)}\bigg]^{1/4}\bigg[\frac{\mathcal{F}(x)}{f_I(x)^4}\bigg]^{1/4}\, ,\nonumber\\
A_I&=-\frac{4 B_I}{\alpha b_I(1+\alpha b_I x)}\dd\phi\, ,\nonumber
\end{align}
with 
\begin{align}
f_{I}(x)&=1+\alpha b_I x\, ,\qquad \mathcal{F}(x)=\prod_{I=1}^{4}f_{I}(x)\, ,\\
h_{I}(y)&=1+\alpha b_I y\, ,\qquad \mathcal{G}(x)=\prod_{I=1}^{4}h_{I}(y)\, ,\nonumber\\
X(x)&=\mathcal{F}(x)\bigg(b_0+\sum_{I=1}^4\frac{16 B_I^2}{f_{I}(x) \alpha^2 b_I\prod_{J\neq I}(b_J-b_I)}\bigg)\, ,\nonumber\\
Y(y)&=\frac{g^2}{\alpha^2}-b_0-\sum_{I=1}^4\frac{16 B_I^2}{h_{I}(y) \alpha^2 b_I\prod_{J\neq I}(b_J-b_I)}\nonumber\, .
\end{align}
The solution depends on 9 parameters: 4 $B_I$, 4 $b_I$ and $\alpha$, whilst $b_0$ can be fixed by a coordinate transformation, this will become important later. We will set the coupling constant $g$ to 1 as in the main text. Imposing supersymmetry will reduce the number of parameters down to 4 in the near-horizon. The form of the solution is reminiscent of the near-horizon solution studied in the main text and it is therefore reasonable that this gives rise to the solution in \eqref{eq:NHmetric}-\eqref{eq:NHscalars}.

\subsection{Near-Horizon limit}

To obtain an AdS$_2$ near-horizon geometry we should find a double root of the function $Y(y)$ and then expand around this point. Since $Y(y)$ is a quartic this is somewhat non-trivial, however in keeping with the results in the main text we do not need to solve for the roots, to take the limit. Instead we will assume the existence of a double root and show that with this assumption we can uniquely fix the near-horizon geometry by ``shooting" for the near-horizon geometry in \eqref{eq:NHmetric}-\eqref{eq:NHscalars}.

Let us fix a double root of $Y(y)$ to be $y_*$, therefore we have
\be
Y(y_*)=0\, ,\quad Y'(y_*)=0\, .
\ee
We can now expand the black hole solution around the horizon. The metric becomes
\be
\dd s^2_4=\frac{\sqrt{\mathcal{F}(x)\mathcal{H}(y_*)}}{\alpha^2(x-y_*)^2}\bigg[-\frac{Y''(y_*)}{2}(y-y_*)^2\dd t^2+ \frac{2\mathcal{H}(y_*)}{Y''(y_*)(y-y_*)^2}\dd y^2+\frac{\dd x^2}{X(x)}+\frac{X(x)}{\mathcal{F}(x)}\dd\phi^2\bigg]\, .
\ee
After a coordinate redefinition it takes the form
\be
\dd s^2_4=\frac{\sqrt{\mathcal{F}(x)\mathcal{H}(y_*)}n^2}{\alpha^2(x-y_*)^2}\bigg[\dd s^2(\text{AdS}_2)+\frac{\dd x^2}{n^2 X(x)}+\frac{m^2X(x)}{n^2\mathcal{F}(x)}\dd z^2\bigg]\, ,
\ee
where
\be
n^2=\frac{2\mathcal{H}(y_*)}{Y''(y_*)}\, ,\quad \phi=m z\, .\label{eq:defn}
\ee
In comparing with \eqref{eq:NHmetric} we identify
\be
\mathcal{F}(x)\mathcal{H}(y_*)n^2=P(w)\alpha^2 (x-y_*)^2\, .
\ee
Next consider the coefficient of $\dd z^2$, comparing with \eqref{eq:NHmetric} we find
\be
X(x)=\frac{(x-y_*)^4\alpha^4 f(w)}{m^2 n^2 \mathcal{H}(y_*)}\, ,
\ee
and by equating the $\dd x^2$ term and the $\dd w^2$ term in \eqref{eq:NHmetric} we find
\be
x=y_* +\frac{m \sqrt{\mathcal{H}(y_*)}}{w \alpha^2}\, .\label{eq:xCOC}
\ee
With these definitions the metric takes the same form as in \eqref{eq:NHmetric}. It remains to check the other fields and that the expressions for $X(x)$ and $\mathcal{F}(x)$ are actually consistent with the change of coordinates in \eqref{eq:xCOC} and the expressions for $f(w)$ and $P(w)$ in the main text. 

Next, the scalars are shown to be equivalent in the near-horizon limit provided
\be
m=\alpha n \, ,\qquad b_I=-\frac{c_I}{\alpha c_I y_* +n \sqrt{\mathcal{H}(y_*)}}\, .\label{eq:newbs}
\ee
The parameters $B_I$ are fixed by studying the gauge fields, and we find
\be
B_I=\frac{c_I \sqrt{\mathcal{H}(y_*)}}{4\big(\alpha c_1 y_*+n \sqrt{\mathcal{H}(y_*)\big)}}\, .\label{eq:Bdef}
\ee
Note that a relation between $B_I$ and $c_I$ is to be expected if the full black hole was supersymmetric and therefore this is quite natural from this point of view. In fact one sees that these relations turn out to be equivalent to $Y(y)$ getting a double root\footnote{One can show that 
\begin{equation*}
Y(y)=\frac{(y-y_*)^2}{n^2 \mathcal{H}(y)}
\end{equation*}
after applying the solution for $B_I$.} , and therefore one should think of this as the extremal condition for the black hole. We therefore have that the near-horizon limit of the black hole solution in \eqref{eq:fullBH} takes the correct general form of the solution studied in the main text. What remains to be checked is that the definitions of $X(x)$ and $\mathcal{F}(x)$ are consistent with the form of $f(w)$ and $P(w)$ in the main text respectively. It turns out that checking $\mathcal{F}(x)$ gives the correct form for $P(w)$ is straightforward. By substituting in the change of coordinates \eqref{eq:xCOC} and \eqref{eq:newbs} one simply lands on the correct form for $P(w)$ and therefore it is consistent, for $X(x)$ and $f(w)$ this is not as simple. 

Firstly, we set $b_0=\frac{1}{\alpha^2}$ and $g=1$ which implies that the first two terms in $Y(y)$ cancel. After a little rewriting we find that $X(x)$ takes the form
\begin{align}
X(x)=\frac{ \mathcal{H}(y_*)}{\alpha^2 w^4}\bigg[ P(w)+w\Big(T_0+T_1 w+T_2 w^2+T_3 w^3\Big)\bigg]\, ,
\end{align}
with $T_i$ some coefficients. At first these coefficients look particularly unwieldy however we may express them in terms of $Y(y_*)$ and its derivatives. We find
\begin{align}
T_0&=Y(y_*)\sum_I\prod_{I\neq J} c_J-\frac{n\sqrt{\mathcal{H}(y_*)}}{\alpha}Y'(y_*)\sum_{1\leq I<J\leq4} c_I c_J + \frac{n^2\mathcal{H}(y_*)}{2\alpha^2}Y''(y_*)\sum_I c_I -\frac{n^3\mathcal{H}(y_*)^{3/2}}{3! \alpha^3}Y'''(y_*)\, ,\nonumber\\
T_1&=-Y(y_*)\sum_{1\leq I<J\leq 4}c_I c_J+\frac{n\sqrt{\mathcal{H}(y_*)}}{\alpha}Y'(y_*)\sum_{I}c_I -\frac{n^2\mathcal{H}(y_*)}{2\alpha^2}Y''(y_*)\, ,\nonumber\\
T_2&=Y(y_*)\sum_I c_I -\frac{n\sqrt{\mathcal{H}(y_*)}}{\alpha}Y'(y_*)\, ,\nonumber\\
T_3 &=-Y(y_*)\, .
\end{align}
From the assumption of a double root we see that $T_2=T_3=0$ immediately and amazingly after using the definition of $B_I$ in \eqref{eq:Bdef} $T_0$ vanishes also. Therefore only $T_1$ remains and is simply given by
\be
T_1=-\frac{\mathcal{H}(y_*)^2}{\alpha^2}\, ,
\ee
after using \eqref{eq:defn}. Setting
\be
\mathcal{H}(y_*)=\alpha\, ,
\ee
which we are inclined to assume is the final constraint from supersymmetry, implies
\be
X(x)=\frac{\mathcal{H}(y_*)}{\alpha^2 w^4} f(w)\, ,
\ee
and therefore all the definitions are consistent and the near-horizon limit agrees on the nose with the solution discussed in the main text.
We conclude that the supersymmetric limit of the asymptotically AdS$_4$ black hole found in \cite{Lu:2014sza} and given in \eqref{eq:fullBH}, gives rise to the AdS$_2$ geometries studied in this work in the near-horizon limit.

%%%%%%%%%%%%%%%%%%%%%%%%%%%%%%%%%

%%~~~~~~~~~~~~~~~~~~~~~~~~~~~~~~~~~~~~~~~~~~~~~~~~~~~~~~~~~~~~~~~~~~
%%							%%~~~~~~~~~~~~~~~~~~~~~~~~~~~~~~~~~~~~~~~~~~~~~~~~~~~~~~~~~~~~~~~~~~

%%%%%%%%%%%%%%%%%%%

%%%%%%%%%%%%%%%%%%%%%%%%%%%%%%%%%%%%%%%%%
%%%%%%%%%%%%%%%%%%%%%%%%%%%%%%%%%%%%%%%%%

\bibliographystyle{JHEP}

\bibliography{twists}
%%%%%%%%%%%%%%%%%%%%%%%%%%%%%%%%%%%%%%%%
%%%%%%%%%%%%%%%%%%%%%%%%%%%%%%%%%%%%%%%%%

%%%%%%%%%%%%%%%%%%%%%%%%%%%%%%%%%%%%
%%%%%%%%%%%%%%%%%%%%%%%%%%%%%%%%%%%%
\end{document}